\documentclass[12pt]{article}

\usepackage{amsmath}
\usepackage{amsthm}
\usepackage{amssymb}
\usepackage{mathrsfs}
\usepackage{bbm}
\usepackage{graphicx,color}

\newcommand{\ie}{{\it i.e.\/} }
\newcommand{\eg}{{\it e.g.\/} }

\newcommand{\Ran}[1]{\mathrm{Ran}\, #1 }
\newcommand{\Ker}[1]{\mathrm{Ker}\, #1 }

\newcommand{\iu}{\mathrm{i}}
\newcommand{\Id}{1}
\newcommand{\alg}[1]{\mathcal{A}(#1)}
\newcommand{\ideal}[1]{\mathcal{I}_{ #1 }}
\newcommand{\str}{^{*}}
\newcommand{\ep}[1]{\mathrm{e}^{#1}}
\newcommand{\hilb}{\mathcal{H}}
\newcommand{\dd}{\mathrm{d}}
\newcommand{\tr}{\mathrm{tr}}
\newcommand{\Tr}{\mathrm{Tr}}

\newcommand{\teop}{\hfill$\square$}

\newtheorem{thm}{Theorem}
\newtheorem{cor}[thm]{Corollary}
\newtheorem{prop}[thm]{Proposition}
\newtheorem{lma}[thm]{Lemma}

\title{Fredholm determinants and the statistics of charge transport}

\begin{document}
\author{ J.~E.~Avron ${}^{(a)}$, S. Bachmann ${}^{(b)}$, G.M. Graf ${}^{(b)}$
and I. Klich ${}^{(c)}$\\
\normalsize\it
${}^{(a)}$ Department of Physics, Technion, 32000 Haifa, Israel\\
\normalsize\it
${}^{(b)}$ Theoretische Physik, ETH-H\"onggerberg, 8093 Z\"urich, Switzerland\\
\normalsize\it
${}^{(c)}$ Condensed Matter Department,
Caltech, MC 114-36, Pasadena, CA 91125, USA}
\maketitle
\begin{abstract} Using operator algebraic methods we
show that the moment generating function of charge transport in a
system with infinitely many non-interacting Fermions is given by a
determinant of a certain operator in the one-particle Hilbert
space. The formula is equivalent to a formula of Levitov and Lesovik
in the finite dimensional case and may be viewed as its regularized form
in general. Our result embodies two tenets often realized in
mesoscopic physics, namely, that the transport properties are
essentially independent of the length of the leads and of the
depth of the Fermi sea.
\end{abstract}

\section{Introduction}


Models of physical systems are often formulated with the help of one
or few parameters which guarantee that whatever one computes is well
defined and finite while, at the same time, are believed not to
affect properties of physical interest. Examples are: The number of
particles in a macroscopic system, and the lattice spacing
(ultraviolet cutoffs) in the study of critical phenomena.

The theory of transport in mesoscopic systems has two such
parameters: The length of the incoming leads that connect to the
system and the depth of the Fermi sea. The independence of the
length of the leads is the statement that well designed
experiments measure the transport properties of the mesoscopic
system and are independent of the measuring circuit. The
independence of the depth of the Fermi sea expresses the
irrelevance for transport of electrons that are buried deep in the
Fermi sea, 
since in most situations they can not be excited above it. In this
sense there is freedom from both the volume and the ultraviolet scale. See
\cite{Sch} for a numerical investigation of these properties.  

One strategy to address this type of behavior is to consider
idealized systems where the parameters are taken to be infinitely
large. The limiting idealized system comes with the
price tag that expressions for physical quantities that are
otherwise guaranteed to be finite, may become ambiguous, formal
and even infinite. The value in worrying about this idealized,
possibly un-physical system, is precisely in that once the
ambiguities and infinities are resolved, they teach us something
important about the finite physical model, namely, that the
parameters used in its formulation, do indeed effectively
disappear from the physical properties. Their role is effectively
reduced to the control of the small differences between the idealized
model and the physical one.

We shall consider a problem of this kind that arises in the
context of modeling the statistics of charge transport from one
reservoir to another. Levitov and Lesovik \cite{Levitov}
wrote a formula for the appropriate generating function in terms
of a certain infinite dimensional determinant. The formula has found a
number of applications to shot and thermal noise in devices like transmission
barriers, cavities, and interfaces. When one wants to
apply this formula to the idealized cases one finds ambiguities
and, as emphasized by Levitov et al. 
\cite{LevitovLeeLesovik, Ivanov, Levitov2}, the determinant
requires proper definition through regularization. We intend to further the
understanding of these points by providing an alternative,
mathematically consistent, form for the determinant. As we
shall see, the ``regularized form'' of the determinant naturally
emerges once the quantum dynamics is formulated on the state space
of the idealized system. 

In the next section we introduce the statistics of charge transport, review
the Levitov-Lesovik determinant, and propose a regularization. In 
Section~\ref{Results} we state the main results. Section~\ref{Proofs} is 
devoted to proofs and begins with a short overview thereof. Finally,
Section~\ref{exa} exemplifies the assumptions made 
in this work.


\section{The Levitov-Lesovik formula and its regularization}

We consider a lead, where independent electrons are evolved over
some time interval and ask about the statistics of the charge
transferred from the left to the right portion of the lead. 
To begin, we recall the result obtained
in \cite{Levitov} and further elaborated in
\cite{LevitovLeeLesovik,Ivanov}. 
We present its derivation and
generalization to finite times along the heuristic lines given in
\cite{Klich}, in the sense that we do as if the one-particle
Hilbert space $\hilb$ were finite-dimensional.

The fermionic Fock space $\mathcal{F}$ over $\hilb$ contains a
distinguished state, the vacuum, with the physical interpretation
of a no particle state. Let $\Tr$, resp. $\tr$, denote the trace
on $\mathcal{F}$, resp. $\hilb$. Let $U$ be the unitary on $\hilb$
representing the time evolution, and $Q$ the projection
corresponding to the right portion of the lead. Their second
quantizations, $\Gamma(U)=\wedge_{i=1}^k U_i$, resp.
$\dd\Gamma(Q)=\sum_{i=1}^k Q_i$ on $k$-particle states, then stand
for the evolution on $\mathcal{F}$, resp. for the charge in that
portion. We suppose that the initial many particle (mixed) state
is of the form
\begin{equation*}
P=Z^{-1}\Gamma(M)
\end{equation*}
for some operator $M\geq0$, where $Z=\Tr\,\Gamma(M)=\det(1+M)$
ensures that $\Tr P=1$. The reduced one-particle
density matrix $N$ is defined by the property that
\begin{equation*}
\tr(AN)=\Tr\left(\dd\Gamma(A)P\right)
\end{equation*}
for any one-particle operator $A$ on $\hilb$. In our case,
$N=M(\Id+M)^{-1}$. This follows from
\begin{align}
\Tr \left(\ep{\iu\lambda\dd\Gamma(A)}P\right)&=\Tr \left(\Gamma(\ep{\iu\lambda A})P\right)=Z^{-1}\Tr\left(\Gamma(\ep{\iu\lambda A}M)\right)=\frac{\det(\Id+\ep{\iu\lambda A}M)}{\det (\Id+M)} \nonumber \\ &=\det(\Id-N+\ep{\iu\lambda A}N)
\label{TrDet}
\end{align}
by taking the derivative at $\lambda=0$. 

In the following, we assume that $M$ and $Q$, and hence
$P$ and $\dd\Gamma(Q)$, commute, which physically means that in
the state defined by $P$, charge in the lead measured by $Q$ is a
good quantum number. Hence
\begin{equation*}
P|\alpha\rangle=\rho_{\alpha}|\alpha\rangle\,,\qquad\dd\Gamma(Q)|\alpha\rangle=
n_{\alpha}|\alpha\rangle\,,
\end{equation*}
for some basis $\{|\alpha\rangle\}$ of $\mathcal{F}$. The
moment generating function for the charge transfer statistics is
\begin{equation*}
\chi(\lambda)=\sum_{n\in\mathbb{Z}}p_n \ep{\iu\lambda n}\,,
\end{equation*}
where $p_n$ is the probability for $n$ electrons being deposited 
into the right portion 
of the lead by the end of the time
interval. It may be computed as a sum over initial and final
states, $\alpha$ resp. $\beta$, with the former weighted according
to their probabilities $\rho_{\alpha}$:
\begin{align}
\label{BaddefChi}
\chi(\lambda)&=\sum_{\alpha,\beta}|\langle\beta|\Gamma(U)|\alpha\rangle|^2\,\rho_{\alpha}\ep{\iu\lambda(n_{\beta}-n_{\alpha})}=\Tr\left(\Gamma(U)\str\ep{\iu\lambda\dd\Gamma(Q)}\Gamma(U)\ep{-\iu\lambda\dd\Gamma(Q)}P\right) \\
&=Z^{-1}\Tr\left(\Gamma(U\str\ep{\iu\lambda Q}U\ep{-\iu\lambda Q}M)\right)=
\det\left(\Id-N+\ep{\iu\lambda U\str Q U}N\ep{-\iu\lambda Q}\right)\,,
\nonumber
\end{align}
where the trace has been computed in the basis $|\alpha\rangle$, with an
identity $\sum|\beta\rangle\langle\beta|=\Id$ absorbed at the
left of $\Gamma(U)$; the last equality is by~(\ref{TrDet}). This
is the Levitov-Lesovik formula:
\begin{equation}
\label{Levitov}
\chi(\lambda)=\det D(\lambda)\,,\qquad D(\lambda)=N'+\ep{\iu\lambda Q_U}N\ep{-\iu\lambda Q}\,,
\end{equation}
with $N'=\Id-N$ and $Q_U=U\str Q U$. Since $Q$ is a projection,
$\ep{2\pi \iu Q}=\ep{2\pi \iu Q_U}=1$ and $D(\lambda )$ is a periodic
function with period $2\pi$. This expresses the integrality of
charge transport.

An example of a state of interest is that of a system at inverse temperature
$\beta$ having one-particle Hamiltonian $H$; it is $P=Z^{-1}\Gamma(M)$ with
$M=\exp(-\beta H)$ and $N=[\Id+\exp(\beta H)]^{-1}$. In the limit
$\beta\to\infty$, $P$ describes the Fermi sea, whence $N$ is the projection
onto the occupied one-particle states.

\begin{figure}[h]
\begin{center}
\input{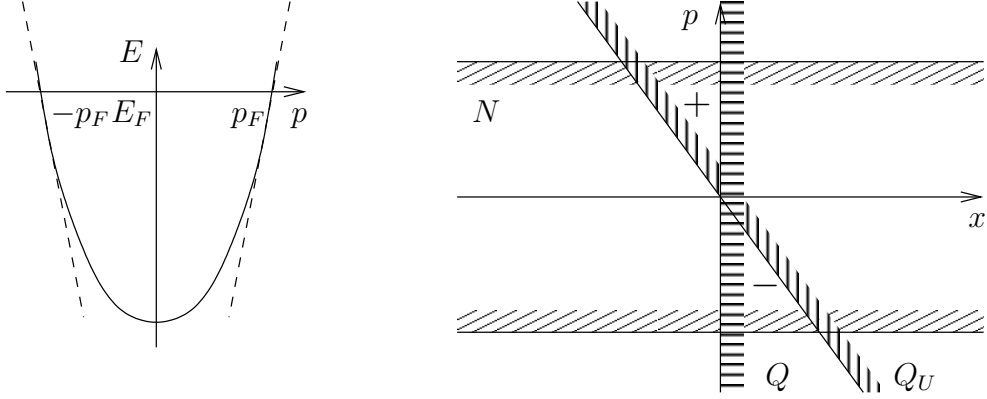}
\caption{Left: dispersion relation $E(p)$ of free particles, and
its linearization. Right: phase space (coordinates $x$, $p$) with
regions selected by $N$, $Q$ and $Q_U$, and hatched along their
boundaries with slanted, horizontal, and vertical dashes, 
respectively.}\label{fig1}
\end{center}
\end{figure}

The above derivation would be rigorous if the one-particle Hilbert space 
were finite dimensional. The question we want to address here is what is the
correct replacement for $D(\lambda)$ when $P$ describes infinitely
many particles, both
because the lead may be infinitely extended spatially (as
appropriate for an open system) and because the Fermi sea may be
very or even infinitely deep. The first concern
appears to affect only the derivation, but not the result,
eq.~(\ref{Levitov}). However, by the second, $D(\lambda)$ differs
from the identity by more than a trace class operator, as would be
required by the definition of a Fredholm determinant. A
manifestation thereof (and in a sense the only one) is that the
expected charge transport
\begin{equation}
\label{ExpTransp}
\langle n \rangle=-\iu\chi'(0)=
-\iu\left.\frac{d}{d\lambda}\det D(\lambda)\right|_{\lambda=0}=
\tr\left((Q_U-Q)N\right)
\end{equation}
involves an operator which is not trace class in the stated situation.
These statements are
illustrated (in the $\beta=\infty$ case) in Fig.~\ref{fig1} representing the
phase space of a single particle moving freely.

The Fermi sea $N$ corresponds to $|p|<p_F$, $p_F$ being the Fermi
momentum, and similarly the right half of the lead $Q$ to $x>0$.
The free evolution, which we take as a simple example for $U$, is
a horizontal shear, so that $Q_U-Q$ is associated with two
sectors, labelled $+$ and $-$. Their intersection with the
horizontal strip associated with $N$ delineates the phase space
support of $(Q_U-Q)N$. Its area, which is a rough estimate of the
trace class norm of the operator, is proportional to the depth of
the sea. If the dispersion relation is conveniently linearized at
$\pm p_F$, the depth becomes infinite, implying that the operator
is not trace class. As a remedy, we note that the expression
\begin{equation*}
\tr\left(QN-Q_UN_U\right)=0\,,
\end{equation*}
with $N_U=U\str N U$, vanishes by splitting the trace, though only 
suggestively so, because the
traces fail to exist separately due to the infinite spatial extent of the
leads. Adding nevertheless that expression to~(\ref{ExpTransp}) yields
\begin{equation}\label{nav}
\langle n \rangle=\tr\left(Q_U(N-N_U)\right)\,,
\end{equation}
which vanishes in the special case of the free evolution, $N_U=N$, and is
expected to be finite in others. This way of
renormalizing the expression is actually declaring that the 
Fermi sea does not
contribute to the current, instead of relying on a compensation between left
and right movers, as indicated by $+$ and $-$ in the figure.

This heuristic manipulation motivates the following regularization of the
Levitov-Lesovik determinant. Replacing $D(\lambda)$ by
\begin{equation}
\label{tildeLevitov}
\widetilde{D}(\lambda)=\ep{-\iu\lambda N_U Q_U}D(\lambda)\ep{\iu\lambda N Q}
\end{equation}
should not change the value of the determinant, since informally
\begin{equation}
\det(\ep{-\iu\lambda N_U Q_U})\cdot\det(\ep{\iu\lambda N
  Q})=\ep{\iu\lambda\,\tr(QN-Q_UN_U)}=1\,.
\label{formal}
\end{equation}
Incidentally, this regularization affects only the first cumulant of the
statistics, \ie the average charge transfer, since the full set of
cumulants is generated
by $\log\det \widetilde{D}(\lambda)$.
We are thus led to recast eq.~(\ref{Levitov}) as
\begin{gather}
\label{nicedet}
\chi(\lambda)=\det\widetilde{D}(\lambda)\,, \\
\label{defnicedet}
\widetilde{D}(\lambda)=\ep{-\iu\lambda N_U Q_U}N'\ep{\iu\lambda N Q}+\ep{\iu\lambda N'_U Q_U}N\ep{-\iu\lambda N' Q}\,.
\end{gather}
It is to be noted that this representation of $\chi(\lambda)$ is
manifestly particle-hole symmetric:
\begin{equation}
\label{PHsymm}
\chi_N(\lambda)=\chi_{N'}(-\lambda)\,.
\end{equation}
It is also $2\pi$-periodic in $\lambda$, though manifestly so only at
$T=0$ since $NQ$, $N'Q$ etc. are all projections.
In that case, eq.~(\ref{defnicedet}) reduces to 
\begin{equation*}
\widetilde{D}(\lambda)=
1+Q_U(N-N_U)\bigl((\ep{\iu\lambda}-1)N-(\ep{-\iu\lambda}-1)N'\bigr)\,,
\end{equation*}
which shows that the generating function $\chi(\lambda)$ is well-defined
whenever its first cumulant (\ref{nav}) is. As we shall see, a 
slightly weaker result holds at positive temperature. 

Let us mention a few connections to other works. A related regularization 
of the Levitov-Lesovik determinant at zero
temperature was used in \cite{MuzykantskyAdamov}, where the
relation of counting statistics to a Riemann-Hilbert problem was
studied. Another one, exhibiting the symmetry (\ref{PHsymm}),
was proposed in \cite{PilgramButtiker}. 
On the more mathematical
side, regularizations of determinants have been related to
renormalization in \cite{ESe}, though by means of a somewhat different 
regularization known as 
$\det_n(1+A)=\det(1+A)\exp{(\tr \sum_{j=1}^{n-1}(-1)^jA^j/j)}$.
The role of C*-algebras in the theory of open systems was recently been
advocated by 
Jak{\v{s}}i{\'c} and Pillet, see \eg \cite{JP}, in general, but also to
fluctuations in particular. A generating function for fluctuations of energy
in bosonic systems has been proposed by \cite{deR}.

The purpose of this work is to
show that, under reasonable assumptions, eq.~(\ref{nicedet})
is obtained {\em without recourse to regularizations}, if the
second quantization is built upon the Fermi sea rather than on the
vacuum $N=0$. 


\section{Results}\label{Results}

Let $\hilb$ be a separable Hilbert space with the following
operators acting on it: An orthogonal projection $Q$, a unitary
$U$, and a selfadjoint $N$, with
\begin{equation}
\label{stateOp}
0\leq N\leq 1\,,
\end{equation}
whose physical interpretations have been described in the previous section. Let
$N'=\Id-N$. We denote by $\ideal{p}$, ($p\ge 1$) the Schatten trace ideals,
\ie the space of all bounded operators $A$ on $\hilb$ such that
$\|A\|_p^p:=\tr|A|^p<\infty$.

The algebra of canonical anticommutation relations (CAR) over $\hilb$ is the
C*-algebra $\alg{\hilb}$ generated by $\Id$, and the elements $a(f)$
and $a\str(f)$, ($f\in\hilb$), such that
\begin{enumerate}
\item the map $f\longmapsto a(f)$ is antilinear
\item $a\str(f)=a(f)\str$
\item these elements satisfy the following anticommutation relations
\begin{equation*}
\{a(f),a\str(g)\}=(f,g)\Id\,,
\end{equation*}
all other anticommutators vanishing.
\end{enumerate}

A (global) gauge transformation is expressed by the automorphism
$\alpha_\lambda: a(f)\mapsto a(\ep{\iu\lambda}f)$. A state
$\omega$ on $\alg{\hilb}$ is gauge-invariant if
$\omega(\alpha_\lambda(A))= \omega(A)$ for all $A\in \alg{\hilb}$.
The operator $N$ defines a gauge-invariant
quasi-free state $\omega_N$ through
\begin{equation}
\omega_N(a\str(f_n)\ldots a\str(f_1)\,a(g_1)\ldots
a(g_m))=\delta_{nm}\det(g_i, N f_j)\,,
\label{state}
\end{equation}
or equivalently by $\omega_N(a\str(f)a(g))=(g, N f)$ and Wick's lemma.
Let $(\hilb_{N}, \pi_N, \Omega_N)$ be the cyclic representation of $\omega_N$:
\begin{equation}
\label{GNS}
\omega_N(A)=(\Omega_N, \pi_N(A)\Omega_N)\,,\qquad (A\in \alg{\hilb})\,.
\end{equation}
The algebra of observables is the (strong) closure of the range of $\pi_N$,
which is equal to its double commutant
$\overline{\pi_N(\alg{\hilb})}=\pi_N(\alg{\hilb})''$. We also recall that a
state is pure if and only if $\pi_N(\alg{\hilb})$ is irreducible, \ie
$\pi_N(\alg{\hilb})'=\left\{c\cdot\Id\mid c\in\mathbb{C}\right\}$, see \eg
\cite{BratRob}, Thm.~2.3.19. This is equivalent to $N$ being a projection 
operator.

These concepts briefly reviewed, we are now ready to state our
main theorem. Its significance is discussed below in a series of
remarks. 
The key result, which is part (v) together with Corollary~\ref{cor},
states that the moment generating function is given by the regularized
determinant, as described in the previous section. 

\begin{thm}
\label{thm1}
Assume that
\begin{gather}
\label{QN}
[Q,N]=0\,, \\
\label{NTrCl}
\sqrt{N}-\sqrt{N_{U\str}}\,,\quad\sqrt{N'}-\sqrt{N'_{U\str}}\in\ideal{1}\,,
\end{gather}
where $N_{U\str}=UN{U\str}$.

\noindent {\rm (Pure state)} Suppose $N=N^2$. Then we have
\begin{enumerate}
\item $\widetilde{D}(\lambda)-\Id\in\ideal{1}$, where
$\widetilde{D}(\lambda)$ is given in eq.~(\ref{defnicedet}).
\item The Bogoliubov automorphisms induced on $\alg{\hilb}$ by the unitary 
operators $U$ and \linebreak $\exp(\iu\lambda Q)$
are implementable on $\hilb_N$: There exist a unitary operator
$\widehat{U}$ and a selfadjoint $\widehat{Q}$ on $\hilb_N$ such that
\begin{align}
\label{UImpl}
\widehat{U}\pi_N\bigl(a^{\#}(f)\bigr)\widehat{U}\str&= \pi_N\bigl(a^{\#}(Uf)\bigr)\,, \\
\label{QImpl}
\ep{\iu\lambda \widehat{Q}}\pi_N\bigl(a^{\#}(f)\bigr)\ep{-\iu\lambda
  \widehat{Q}}&= \pi_N\bigl(a^{\#}(\ep{\iu\lambda Q}f)\bigr)\,,
\end{align}
for all $f\in\hilb$.
\item $\ep{\iu\lambda \widehat{Q}}\in\pi_N\bigl(\alg{\hilb}\bigr)''$.
More
generally, $f(\widehat{Q})\in\pi_N\bigl(\alg{\hilb}\bigr)''$ for any bounded
function $f$.
\item The above properties define $\widehat{U}$ uniquely up to left multiplication with an element from
$\pi_N\bigl(\alg{\hilb}\bigr)'$, and $\widehat{Q}$ up to an additive
constant.
In particular, $\widehat{U}\str \ep{\iu\lambda
\widehat{Q}}\widehat{U} \ep{-\iu\lambda \widehat{Q}}$ is unaffected
by the ambiguities.
\item
\begin{equation}
\label{Main}
(\Omega_N,\widehat{U}\str\ep{\iu\lambda \widehat{Q}}\widehat{U} \ep{-\iu\lambda \widehat{Q}}\Omega_N)=\det \widetilde{D}(\lambda)\,.
\end{equation}
\end{enumerate}

\noindent {\rm (Mixed state)} The above conclusions hold
also for $0<N<\Id$ if, in addition,
\begin{equation}
\label{mixedcond}
Q\sqrt{NN'}\in\ideal{1}\,.
\end{equation}
\end{thm}

\noindent
{\bf Remarks.}
1)  Eq.~(\ref{NTrCl}) demands that the evolution $U$ preserves $N$,
    except for
   creating excitations within an essentially finite region in space and
   energy, as can be seen from the phase space picture given in the
   introduction. This assumption is appropriate for the evolution induced by a
   compact device operating smoothly during a finite time interval.

2)  The operators $\widehat{U}$, $\widehat{Q}$ in (ii) are
    replacements for the
   non-existent $\Gamma(U)$ and $\dd\Gamma(Q)$ mentioned in the
   introduction. Eqs.~(\ref{UImpl}, \ref{QImpl}) state that any additional
   particle in the system evolves by $U$, resp. contributes to the 
   charge as described by $Q$. 

3) If the state is pure, the pair of equations~(\ref{NTrCl}) reduce
to the first one with square roots dropped, and property (iii) holds
trivially, since $\mathcal{B}(\hilb_N)=\pi_N(\alg{\hilb})''$,
the bounded operators on $\hilb_N$. Moreover, $\widehat{U}$ is
unique up to a phase. Incidentally, condition~(\ref{mixedcond})
would be trivial in this case.

4) Property (iii) states that $\widehat{Q}$ is an observable, and
the same is true for $\widehat{U}\str\widehat{Q}\widehat{U}$,
because of $\widehat{U}\str\pi_N(\alg{\hilb})\widehat{U}\subset
\pi_N(\alg{\hilb})$, see~(\ref{UImpl}). Thus, the total charges
before and after the evolution are separately bestowed with physical
meaning.

5) The physical origin of the extra assumption~(\ref{mixedcond})
needed in the mixed state case is as follows. In both cases, the
expected charge contained in a portion of the lead is of order of
its length $L$, or zero if renormalized by subtraction of a
background charge. In the pure case however, the Fermi sea is an
eigenvector of the charge operator, while for the mixed state, the
variance of the charge must itself be of order $L$, because the
occupation of the one-particle states is fluctuating, since
$NN'\neq0$. Hence, in this latter situation, the measurement of the
renormalized charge yields finite values only as long as $L$ is
finite, of which eq.~(\ref{mixedcond}) is a mathematical
abstraction. This condition, while unnecessary for property (ii), is
essential for (iii). Without the latter, the l.h.s. of
eq.~(\ref{Main}) appears to be ambiguous. On the other hand,
the weaker condition
\begin{equation}
(Q_U-Q)\sqrt{NN'}\in\ideal{1}\,,
\label{weaker}
\end{equation}
is sufficient for property (i) and to ensure that the difference
$\widehat{Q_U-Q}$ is an observable.

6) The theorem does not apply to the general
case~(\ref{stateOp}). The two
  cases considered suffice for thermal states with $\beta=\infty$ and
  $0<\beta<\infty$.\\

Let $\widehat{Q}=\int n\, dP(n)$ be the spectral representation of
$\widehat{Q}$. According to quantum mechanical principles an ideal 
measurement of $\widehat{Q}$ with outcome $n$ in $dn$ collapses 
$\Omega_N$ to the state $dP(n)\Omega_N$, normalized to the probability 
$(dP(n)\Omega_N, dP(n)\Omega_N)$ of that outcome. Effectively, this means 
that $dP(n)\Omega_N$ is the state relevant for a second measurement. The
charge transfer is inferred from two measurements \cite{MuzykantskyAdamov} of
the charge $\widehat{Q}$, one before and one after the evolution of the 
system by 
$\widehat{U}$. The joint probability for measurements $n$ and $m$ is 
$(\widehat{U}dP(n)\Omega_N,dP(m) \widehat{U}dP(n)\Omega_N)$ and the
generating function appropriately defined as
\begin{equation*}
\chi_N(\lambda)=\iint(dP(n)\Omega_N, \widehat{U}^*dP(m)\widehat{U}dP(n)\Omega_N)
\ep{\iu\lambda(m-n)}\,.
\end{equation*}
\begin{cor}\label{cor}
The spectrum of $\widehat{Q}$ consists of integers, up to an additive 
constant. 
The generating function is
\begin{equation*}
\chi_N(\lambda)=(\Omega_N, \widehat{U}\str\ep{\iu\lambda\widehat{Q}}\widehat{U}\ep{-\iu\lambda\widehat{Q}}\Omega_N)
\end{equation*}
and describes the transport of integer charges $n$ with non-negative
probabilities: 
\begin{equation*}
\chi_N(\lambda)=\sum_{n\in\mathbb{Z}}p_n\ep{\iu\lambda n}\,,\qquad
p_n\geq0 \,,\quad\sum_{n\in\mathbb{Z}}p_n=1\,.
\end{equation*}
Moreover, the particle-hole symmetry~(\ref{PHsymm}) holds true.
\end{cor}


\section{Proofs}\label{Proofs}
We begin by giving the proof of the corollary and continue with that of part
(i) of the theorem. Then we give some preliminaries, including 
details such as
inner Bogoliubov automorphisms and the Shale-Stinespring criterion for 
general ones. Thereafter we prove parts (ii-iv) readily if the state is pure, 
and using its purification, if it is mixed. Finally, the main formula (v) is
obtained using an approximation procedure in terms of inner automorphisms and 
finite dimensional determinants. 

\subsection{Proof of Corollary \ref{cor}}
We begin by recalling that every gauge-invariant state is a factor state (see
\cite{PS}, Thm.~5.1), \ie
\begin{equation}
\label{factor}
\pi_N(\alg{\hilb})'\cap\pi_N(\alg{\hilb})''=\{c\cdot 1 |\,c\in\mathbb{C}\}\,.
\end{equation}
{From} eq.~(\ref{QImpl}) and $\ep{2\pi\iu Q}=1$, we see that
$\ep{2\pi\iu \widehat{Q}}\in\pi_N(\alg{\hilb})'$, while by~(iii)
we have $\ep{2\pi\iu \widehat{Q}}\in\pi_N(\alg{\hilb})''$.
Thus $\ep{2\pi\iu \widehat{Q}}=c$, $(|c|=1)$ and we may assume
$c=1$ by adding an additive constant to $\widehat{Q}$, see~(iv).
The spectral representation of $\widehat{Q}$ is then of the form
\begin{equation}
\label{QSpectrDec}
\widehat{Q}=\sum_{n\in\mathbb{Z}} nP_n\,.
\end{equation}
We note that
\begin{equation}
\label{someExp}
(\Omega_N, \ep{\iu\lambda\widehat{Q}}A\ep{-\iu\lambda\widehat{Q}}\,\Omega_N)=
(\Omega_N, A\,\Omega_N)
\end{equation}
for $A\in\pi_N(\alg{\hilb})''$. Indeed, for $A=\pi_N(a\str(f)\,a(g))$, we have
$\ep{\iu\lambda\widehat{Q}}A\ep{-\iu\lambda\widehat{Q}}=
\pi_N(a\str(\ep{\iu\lambda Q}f)$ $a(\ep{\iu\lambda Q}g))$ by~(\ref{QImpl}).
The expectations~(\ref{someExp}) agree because of
$(\ep{\iu\lambda Q} g, N \ep{\iu\lambda Q} f)=(g, N f)$
by $[Q,N]=0$. The same holds true by~(\ref{state}) for arbitrary products of
$a\str(f_i)$, $a(g_i)$, and by density, for
$A\in\pi_N(\alg{\hilb})''$. By~(iii) we may apply~(\ref{someExp}) to
$A\ep{\iu\lambda\widehat{Q}}$ instead of $A$, and obtain $(\Omega_N,
P_n A\,\Omega_N)=(\Omega_N, AP_n\,\Omega_N)$; then this to $A P_n \in\pi_N(\alg{\hilb})''$ instead of $A$, and get $(\Omega_N,
A P_n\,\Omega_N)=(\Omega_N, P_n A P_n\,\Omega_N)$. Moreover, we have
$\widehat{U}\str\ep{\iu\lambda\widehat{Q}}\widehat{U}\in\widehat{U}\str\pi_N(\alg{\hilb})''\widehat{U}\subset \pi_N(\alg{\hilb})''$
by~(\ref{UImpl}). Hence, using~(\ref{QSpectrDec}), we see that
\begin{align*}
(\Omega_N, \widehat{U}\str\ep{\iu\lambda\widehat{Q}}\widehat{U}\ep{-\iu\lambda\widehat{Q}}\,\Omega_N)&=\sum_{n\in\mathbb{Z}}\,(\Omega_N,
P_n\widehat{U}\str\ep{\iu\lambda\widehat{Q}}\widehat{U}P_n\,\Omega_N)\ep{-\iu\lambda n} \\
&=\sum_{n,m\in\mathbb{Z}}\,(\Omega_N,P_n\widehat{U}\str P_m\widehat{U}P_n\,\Omega_N)\ep{\iu\lambda (m-n)}
\end{align*}
is of the stated form.\teop\newline

\subsection{Part (i)}

Since the projection $Q$ commutes with $N$, see (\ref{QN}), we have
\begin{align*}
\ep{\iu\lambda NQ}&=\Id+(\ep{\iu\lambda N}-\Id)Q\,, \\
\ep{-\iu\lambda N_UQ_U}&=\Id+Q_U(\ep{-\iu\lambda N_U}-\Id)\,.
\end{align*}
We insert these equations in the definition (\ref{defnicedet}) of
$\widetilde{D}(\lambda)$. Moreover,
\begin{equation}
N_U-N=N^{1/2}(N_U^{1/2}-N^{1/2})+(N_U^{1/2}-N^{1/2})N_U^{1/2}\in\ideal{1}\,,
\label{deltaN}
\end{equation}
so that
\begin{equation*}
\ep{-\iu\lambda N}-\ep{-\iu\lambda N_U}=
\iu\int_0^\lambda\ep{-\iu(\lambda-s)N_U}(N_U-N)\ep{-\iu s N}\,\dd s
\end{equation*}
also belongs to the trace class ideal. Rather than proving
$\widetilde{D}(\lambda)\in\Id+\ideal{1}$ for
$\widetilde{D}(\lambda)$ we may thus do so for the expression
\begin{multline*}
[\Id+Q_U(\ep{-\iu\lambda N}-\Id)]N'[\Id+(\ep{\iu\lambda N}-\Id)Q]+
[\Id+Q_U(\ep{\iu\lambda N'}-\Id)]N[\Id+(\ep{-\iu\lambda N'}-\Id)Q]\\
\begin{aligned}
=&N'+N&&+Q_U[(\ep{-\iu\lambda N}-\Id)N'+(\ep{\iu\lambda N'}-\Id)N]+[N'(\ep{\iu\lambda N}-\Id)+N(\ep{-\iu\lambda N'}-\Id)]Q \\
&&&+Q_U[(\ep{-\iu\lambda N}-\Id)N'(\ep{\iu\lambda N}-\Id)+(\ep{\iu\lambda N'}-\Id)N(\ep{-\iu\lambda N'}-\Id)]Q \\
=&\Id&&+Q_U\left[(\cos(\lambda N)-\Id)N'+(\cos(\lambda N')-\Id)N-\iu\sin(\lambda N) N'+\iu\sin(\lambda N') N\right] \\
&&&+Q\left[(\cos(\lambda N)-\Id)N'+(\cos(\lambda N')-\Id)N+\iu\sin(\lambda N) N'-\iu\sin(\lambda N') N\right] \\
&&&+2Q_UQ\left[(\Id-\cos(\lambda N))N'+(\Id-\cos(\lambda N'))N\right] \\
=&\Id&&+(Q_U^2+Q^2-2Q_UQ)[(\cos(\lambda N)-\Id)N'+(\cos(\lambda N')-\Id)N] \\
&&&+\iu(Q-Q_U)[\sin(\lambda N)N'-\sin(\lambda N')N]\,.
\end{aligned}
\end{multline*}
With the help of the functions $f(x)=(\cos x-1)/x$ and
$g(x)=(\sin x)/x$, which are bounded also at $x=0$, the expression is
rewritten as 
\begin{equation*}
\Id+[(Q-Q_U)Q-Q_U(Q-Q_U)]NN'\lambda(f(\lambda N)+f(\lambda N'))+
\iu(Q-Q_U)NN'\lambda(g(\lambda N)-g(\lambda N'))\,.
\end{equation*}
Besides of $Q\sqrt{NN'}\in\ideal{1}$, see eq.~(\ref{mixedcond}),
we have $Q_U\sqrt{NN'}=U\str
Q\sqrt{N_{U\str}N'_{U\str}}U\in\ideal{1}$ by eq.~(\ref{NTrCl}),
and hence $(Q-Q_U)\sqrt{NN'}\in\ideal{1}$, cf.~(\ref{weaker}).
This makes the claim manifest.
\teop\newline

In the zero temperature case, where $N$ is
a projection, the above proof simplifies considerably due to $NN'=0$.

\subsection{Preliminaries}\label{pre}

We recall a few results about Bogoliubov transformations, first inner and then
others.

Given a bounded operator $A$ on $\hilb$, operators $\Gamma(A)$ and
$\dd\Gamma(A)$ are usually defined on the Fock space over $\hilb$. Following
\cite{ArakiWyss} we
define them instead as elements of the CAR-algebra $\alg{\hilb}$, when $A$ is
of finite rank.

\begin{itemize}
\item For rank one operators $A_i=|f_i\rangle\langle g_i|$, $(i=1,\ldots,n)$,
we set
\begin{equation}
\label{DefDGamma}
\dd\Gamma(A_1,\,\ldots,\,A_n)=
a\str(f_n)\cdots a\str(f_1)\,a(g_1)\cdots a(g_n)\,.
\end{equation}
The definition is extended by multilinearity to operators $A_i$ of finite
rank. The result is independent of the particular decomposition into rank
one operators.
\item For $U-\Id$ of finite rank, we set
\begin{equation*}
\Gamma(U)=\sum_{n=0}^\infty\frac{1}{n!}\dd\Gamma(\underbrace{U-\Id,\,\ldots,\, U-\Id}_{n})\,,
\end{equation*}
where the term $n=0$, in which no arguments are present, is read as
$\dd\Gamma=1$. The sum is finite, because the terms with
$k>\mathrm{rank}(U-1)$ vanish.
\end{itemize}
The elements of $\alg{\hilb}$ just defined share the properties of the
operators on Fock space known by the same notation.

\begin{lma}
Let $U-1$ be of finite rank. Then
\begin{align}
\Gamma(U)a\str(f)&=a\str(Uf)\Gamma(U)\,, \label{prp1}\\
\Gamma(U_1U_2)&=\Gamma(U_1)\Gamma(U_2)\,.\label{prp3}
\end{align}
In particular, $\Gamma(U)$ is unitary if $U$ is.
\end{lma}
\paragraph{Proof.} We have
\begin{equation}
\label{dga}
\dd\Gamma(A_1,\ldots,A_n)a\str(f)=a\str(f)\dd\Gamma(A_1,\ldots,A_n)+\sum_{i=1}^{n}\,a\str(A_if)\dd\Gamma(A_1,\ldots,\widehat{A}_i,\ldots,A_n)\,
\end{equation}
where the hat indicates omission. In the rank one case,
$A_i=|f_i\rangle\langle g_i|$, this follows
from~(\ref{DefDGamma}) and from $(g_i,f)a\str(f_i)=a\str(A_i f)$. In the
general case, by multilinearity. Thus,
\begin{align*}
\Gamma(U)a\str(f)&=a\str(f)\Gamma(U)+a\str((U-1)f)\sum_{n=1}^{\infty}\frac{1}{(n-1)!}\,\dd\Gamma(U-1,\ldots,U-1) \\
&=a\str(f)\Gamma(U)+a\str((U-1)f)\Gamma(U) = a\str(Uf)\Gamma(U)\,,
\end{align*}
since we applied (\ref{dga}) with $n$ equal entries $A_i=U-1$.

\noindent We have
\begin{align*}
&\dd\Gamma(A_1,\ldots,A_n)\dd\Gamma(B_1,\ldots,B_m)= \\
&=\sum_{l=0}^{\min(n,m)}\sum_{\mathcal{C}_l}\,\dd\Gamma(A_{i_1}B_{j_1},\ldots,A_{i_l}B_{j_l},A_1,\ldots,\widehat{A}_{i_s},\ldots,A_n,B_1,\ldots,\widehat{B}_{j_s},\ldots,B_m)\,,
\end{align*}
where the second sum runs over all $l$-contractions $(i_1, j_1),\ldots,(i_l, j_l)$ with $i_1<\ldots<i_l,j_{i_s}\neq j_{i_r}$. In the rank one
case, which implies the general one, this is just Wick's lemma for normal
ordered products. Thus
\begin{align*}
\Gamma(U_1)&\Gamma(U_2)=\sum_{n=0}^{\infty}\frac{1}{n!}\,\dd\Gamma(U_1-1,\ldots,U_1-1)\cdot\sum_{m=0}^{\infty}\frac{1}{m!}\,\dd\Gamma(U_2-1,\ldots,U_2-1)
\\
&=\sum_{n,m=0}^{\infty}\sum_{l=0}^{\min(n,m)}\frac{1}{l!(n-l)!(m-l)!}\,\dd\Gamma((U_1-1)(U_2-1),\ldots,U_1-1,\ldots,U_2-1,\ldots)
\end{align*}
with entries repeated $l$, $n-l$, $m-l$ times. In fact, the number of $l$-contractions is
\begin{equation*}
\frac{1}{l!}\frac{n!}{(n-l)!}\frac{m!}{(m-l)!}\,.
\end{equation*}
Setting $n-l=:s$, $m-l=:t$, $l+s+t=:r$, we have
\begin{align*}
\Gamma(U_1)\Gamma(U_2)&=\sum_{r=0}^{\infty}\sum_{\begin{subarray}{c} l,s,t \\
l+s+t=r\end{subarray}}\frac{1}{l!\,s!\,t!}\,\dd\Gamma((U_1-1)(U_2-1),\ldots,U_1-1,\ldots,U_2-1,\ldots) \\
&=\sum_{r=0}^{\infty}\frac{1}{r!}\,\dd\Gamma((U_1-1)(U_2-1)+(U_1-1)+(U_2-1),\ldots)
\end{align*}
since there are ${r!}/l!\,s!\,t!$ ways to pick terms from each entry of
the last line. Since $(U_1-1)(U_2-1)+(U_1-1)+(U_2-1)=U_1U_2-1$,
the proof is complete.\teop\newline

If $O$ is an operator on $\hilb$ such that $O-\Id$ is in the trace
class, its {\rm Fredholm determinant} is defined by
\begin{equation}
\label{FredDet}
\det O=\sum_{k=0}^{\infty}\tr\wedge^k(O-\Id)\,.
\end{equation}
This extends the usual definition of the determinant in the finite
dimensional case.
\begin{lma}
\label{scalprod}
Let $A$ be a finite rank operator, and $0\leq N\leq\Id$. Then
\begin{equation*}
\omega_N(\dd\Gamma(\underbrace{A,\,\ldots,\,A}_k))=\tr\wedge^k(AN)\,.
\end{equation*}
Moreover, if $U$ is such that $U-\Id$ is of finite rank, then
\begin{equation}
\omega_N(\Gamma(U))=\det((\Id-N)+UN)\,.
\label{det}
\end{equation}
\end{lma}
\paragraph{Proof.} The trace of a finite rank operator
$A=\sum_{i=1}^m f_i(g_i, \cdot)$ is $\tr A=\sum_{i=1}^m (g_i,f_i)$. By the same
token, that of
\begin{equation*}
\wedge^kA=\sum_{i_1,\ldots i_k=1}^m \frac{1}{k!}\sum_{\sigma\in S_k}
(-1)^\sigma\otimes_{\alpha=1}^kf_{i_{\sigma(\alpha)}}(g_{i_\alpha},  \cdot)
\end{equation*}
is
\begin{equation*}
\tr\wedge^kA=
\sum_{1\leq i_{1}<\ldots<i_{k}\leq m}
\det(g_{i_{\alpha}}, f_{i_{\beta}})_{ \alpha, \beta=1}^k\,.
\end{equation*}
Since the $a^{\#}(f)$ anticommute, we have
\begin{equation*}
\dd\Gamma(A,\,\ldots,\,A)=
\sum_{1\leq i_{1}<\ldots<i_{k}\leq m}a\str(f_{i_1})\cdots a\str(f_{i_k})\,a(g_{i_k})\cdots a(g_{i_1})\,,
\end{equation*}
whose expectation value is computed by (\ref{state}) as
\begin{equation*}
\omega_N(\dd\Gamma(A,\,\ldots,\,A))
=\sum_{1\leq i_{1}<\ldots<i_{k}\leq m}
\det(Ng_{i_{\alpha}}, f_{i_{\beta}})_{ \alpha, \beta=1}^k =
\tr\wedge^k(AN) \,,
\end{equation*}
because the decomposition of $AN$ differs from that of $A$ by
$Ng_i$ in place of $g_i$. This proves the first part of the lemma. The second
part is now an application of the definition of the determinant,
eq.~(\ref{FredDet}). Indeed,
$\omega_N(\Gamma(U))=1+\sum_{k=1}^{\infty}\tr\wedge^k((U-\Id)N)=\det(\Id+(U-\Id)N)$.\teop\newline

We recall a few results on Bogoliubov
transformations. Their proofs can be found \eg in \cite{Araki_2},
where however CAR-algebras are introduced in the self-dual guise.
A remark at the end of this subsection is intended as an aid to
translation. The first result is the Shale-Stinespring 
criterion \cite{ShSt} about unitary implementability (see \eg \cite{Lundberg},
\cite{Araki_2} Thm. 6.3 (1)).
\begin{prop}
\label{BogImpl_1} Let $P$ be a projection and $V$ a unitary
operator on $\hilb$. The Bogoliubov automorphism induced by $V$ on
$\hilb$, \ie $a(f)\mapsto a(Vf)$, is unitarily implementable in
the representation $\pi_P$ if and only if $PV(\Id-P)$ and
$(\Id-P)VP$ are in the Hilbert-Schmidt class, $\ideal{2}$.
\end{prop}
In particular, an equivalent condition is
$[P, V]=PV(\Id-P)-(\Id-P)VP\in\ideal{2}$. There is a version of this
proposition for groups (\cite{Araki_2}, Thm. 6.10 (2, 3)).
\begin{prop}
\label{BogImpl_2}
Let $V$ in the previous proposition be replaced by a 1-parameter unitary group
$V_\lambda$, ($\lambda\in\mathbb{R}$), such that $V_\lambda$ is norm
continuous and $PV_\lambda(\Id-P)$ is continuous in the
$\ideal{2}$-norm. Then $V_\lambda$ has an implementer of the form
$\widehat{V}_\lambda=\exp(\iu\lambda \widehat{v})$, where $\widehat{v}$ is a
self-adjoint operator on $\hilb_{P}$. The requirement
\begin{equation}
\label{nrmlz}
(\Omega_P,\widehat{v}\Omega_P)=0
\end{equation}
may be imposed, in which case $\widehat{v}$ is unique.
\end{prop}
An equivalent condition,
is $V_\lambda=\exp(\iu\lambda v)$ with $v$ a bounded, selfadjoint
operator on $\hilb$, and $Pv(\Id-P)\in\ideal{2}$.

The next result is about the
continuity of the implementation, see \cite{Araki_2}, Thm.~6.10 (7).
\begin{prop}
\label{continuity} Let $v$ and $v_n$, ($n=0,1,\ldots$), satisfy the hypotheses
of Prop.~\ref{BogImpl_2}, and let $\widehat{v}$, $\widehat{v_n}$ satisfy the
normalization (\ref{nrmlz}). If $(v_n)$ converge strongly to $v$ and if
$\| P(v_n-v)(\Id-P)\|_2\rightarrow 0$ as $n\to\infty$, then
\begin{equation*}
\mathrm{s-}\lim_{n}\ep{\iu\lambda \widehat{v_n}}=\ep{\iu\lambda \widehat{v}}\,.
\end{equation*}
\end{prop}

The last preliminary is concerned with the twisted duality of CAR-algebras.
Let $P$ be an orthogonal projection on $\hilb$,
$\mathcal{K}\subset\hilb$ a closed subspace, and
$\mathcal{K}^{\perp}$ its orthogonal complement. Let
$\widetilde{\mathcal{A}}({\mathcal{K}^\perp})$ be the von Neumann algebra
generated by $\widehat{\Lambda}\pi_P(a(f))$, ($f\in\mathcal{K}^\perp$), where
$\widehat{\Lambda}$ is the parity. Then \cite{HaagDuality, CARDuality}
\begin{equation}
\label{duality}
\pi_P(\alg{\mathcal{K}})'=\widetilde{\mathcal{A}}({\mathcal{K}^{\perp}})\,.
\end{equation}
The implementers $\widehat{V}$ from Prop.~\ref{BogImpl_1} commute with
parity:
Let $\widehat{\Lambda}:\:\hilb_P\rightarrow\hilb_P$ be the unitary
implementation of the *-automorphism $a(f)\mapsto a(-f)$ which is uniquely
determined by $\widehat{\Lambda}\Omega_P=\Omega_P$. Then
\begin{equation}
[\widehat{V},\widehat{\Lambda}]=0\,,
\label{comm}
\end{equation}
see \cite{Araki_2}, Thm. 6.3 (3), Thm. 6.7 (2). We will actually apply this
fact only to $\widehat{V}=\widehat{V}_\lambda$ as in Prop.~\ref{BogImpl_2}, in
which case it can be verified as follows. Since $[V_\lambda, -1]=0$, the
operators $\widehat{\Lambda}\widehat{V}_\lambda$ and
$\widehat{V}_\lambda\widehat{\Lambda}$ implement the same Bogoliubov
automorphism, whence
$\widehat{\Lambda}\widehat{V}_\lambda=
c_\lambda\widehat{V}_\lambda\widehat{\Lambda}$ with $|c_\lambda|=1$. From
$(\Omega_P, \widehat{V}_\lambda\Omega_P)=
(\widehat{\Lambda}\Omega_P, \widehat{\Lambda}\widehat{V}_\lambda\Omega_P)
=c_\lambda(\Omega_P, \widehat{V}_\lambda\Omega_P)$ we find $c_\lambda=1$
for small $\lambda$, because $(\Omega_P, \widehat{V}_\lambda\Omega_P)\to 1$,
($\lambda\to 0$). The conclusion extends to all $\lambda$ by the group
property.\\

\noindent
{\bf Remark.} In order to make contact with the repeatedly cited
article \cite{Araki_2} we recall that a self-dual CAR-algebra
$\alg{\widetilde\hilb,\Gamma}$ is given in terms of a separable Hilbert space
$\widetilde\hilb$ equipped with a conjugation $\Gamma$. Its
generators $B(h)$ are linear in $h\in\widetilde\hilb$ and the relations are
$B(h)^*=B(\Gamma h)$ and $\{B(h),B(h')^*\}=(h',h)1$.
A projection $\widetilde P$ on $\widetilde\hilb$ satisfying
$\widetilde P+\Gamma\widetilde P\Gamma=1$ defines a pure state
$\omega$ on the algebra through
\[
\omega(B(h)^*B(h))=0\,, \qquad (\widetilde Ph=0)\,.
\]
The algebra $\alg{\hilb}$ is connected to the above by picking a
conjugation $C$ on $\hilb$ and by setting
\[
\widetilde\hilb=\hilb\oplus \hilb\,,\qquad
\Gamma (f\oplus g)=Cg\oplus Cf\,,\qquad
B(f\oplus g)=a^*(f)+a(Cg)\,.
\]
States defined by $P$ and $\widetilde P$ then agree if
$\widetilde P(f\oplus g)=(1-P)f\oplus CPCg$.
\subsection{Parts (ii-iv)}

{\bf Pure state:} $N=N^2$

Existence: In the case $V_\lambda=\exp(\iu\lambda Q)$ we have
$[N, V_\lambda]=0$ by eq.~(\ref{QN}), so that existence of a unitary
implementer $\widehat{V}_\lambda=\exp(\iu\lambda \widehat{Q})$ is trivial by
Prop.~\ref{BogImpl_2}. Similarly, in the case $V=U$ we have
$[N,U]=(N-N_{U\str})U\in\ideal{2}$ by eq.~(\ref{deltaN}). Hence it is also
implementable by Prop.~\ref{BogImpl_1}.

Uniqueness:
Let $\widehat{V}$ denote either of $\exp(\iu\lambda \widehat{Q})$ or
$\widehat{U}$. Suppose $\widehat{V}_1$ and $\widehat{V}_2$ both implement the
same transformation. Then
$\widehat{V}_1\widehat{V}_2\str\pi_N(a(Vf))=
\pi_N(a(Vf))\widehat{V}_1\widehat{V}_2\str$.
Thus $\widehat{V}_1=(\widehat{V}_1\widehat{V}_2\str)\widehat{V}_2$ and
$\widehat{V}_2$ differ by left multiplication with
$\widehat{V}_1\widehat{V}_2\str\in \pi_N(\alg{\hilb})'$. In the pure case the
cyclic representation is irreducible, whence $\widehat{U}$ is unique up to a
phase and $\widehat{Q}$ up to an additive constant. As mentioned in Remark 3,
property (iii) is empty in this case.\\

\noindent{\bf Mixed state:} $0<N<\Id$

Given $0\leq N\leq\Id$ on $\hilb$, we consider its {\it purification}
\begin{equation*}
P_N=\left(\begin{array}{cc} N & \sqrt{NN'} \\ \sqrt{NN'} &
    N'\end{array}\right)
=P_N^2
\end{equation*}
on $\hilb\oplus\hilb$, together with the cyclic (or GNS) representation
$(\hilb_{P_N}, \pi_{P_N}, \Omega_{P_N})$ of the state defined by $P_N$ on $\alg{\hilb\oplus\hilb}$. We
can identify
\begin{equation*}
\alg{\hilb}\cong\alg{\hilb\oplus0}
\end{equation*}
via $a(f)=a(f\oplus0)$, and
\begin{equation}
\label{equiv}
\hilb_N\equiv\pi_{P_N}(\alg{\hilb\oplus0})''\Omega_{P_N}\subset\hilb_{P_N}\,,
\qquad
\Omega_N\equiv\Omega_{P_N}\,,\qquad
\pi_N(a)\equiv\pi_{P_N}(a)\upharpoonright\hilb_N\,,
\end{equation}
since these objects satisfy
\begin{equation}
(\Omega_{P_N},\pi_{P_N}(a\str(f\oplus0)a(g\oplus0))\Omega_{P_N})
=(g\oplus0, P_N(f\oplus0))=(g, Nf)\,,
\end{equation}
as required by (\ref{GNS}). We can not handle the most general mixed case
$0\le N\le 1$. The reason comes from the following
lemma, whose proof is postponed till the end of the section.

\begin{lma}\label{cyclic}
Assume $0<N<1$ (strict inequality). Then $\,\Omega_{P_N}$ is cyclic in
$\hilb_{P_N}$ for $\pi_{P_N}(\alg{\hilb\oplus0})$. In particular, we have
equality in~(\ref{equiv}), $\hilb_N=\hilb_{P_N}$.
\end{lma}

 A unitary $V$ on $\hilb$ induces two automorphisms on
$\alg{\hilb\oplus\hilb}$: (a) $a(f\oplus g)\mapsto a(Vf\oplus g)$,
and (b) $a(f\oplus g)\mapsto a(Vf\oplus Vg)$, whose implementation
may be envisaged:
\begin{align}
\text{(a)}\qquad
\widehat{V}\pi_{P_N}(a(f\oplus g))\widehat{V}\str
=&\pi_{P_N}(a(Vf\oplus g))\,,\label{impla} \\
\text{(b)}\qquad
\widehat{V}\pi_{P_N}(a(f\oplus g))\widehat{V}\str
=&\pi_{P_N}(a(Vf\oplus Vg))\,. \nonumber
\end{align}
Both choices for $\widehat{V}$ would provide an implementation for $V$ in
the representation $\pi_N$ on
$\hilb_N$. Only in the first case we have
$\widehat{V}\in\pi_N(\alg{\hilb})''$. Indeed, by (\ref{duality}) applied to
$\hilb\oplus 0\subset \hilb\oplus\hilb$ instead of $\mathcal{K}\subset\hilb$,
we need to check $\widehat{V}\in \widetilde{\mathcal{A}}(0\oplus\hilb)'$.
This however follows from $[\widehat{V}, \pi_{P_N}(a(0\oplus g))]=0$, see
(\ref{impla}), and from $[\widehat{V},\widehat{\Lambda}]=0$, see
(\ref{comm}).

In order to determine the existence of these implementations we compute
\begin{equation*}
\left[P_N, \left(\begin{array}{cc} U_1 & 0 \\ 0 & U_2 \end{array}\right)\right]=\left(\begin{array}{cc} [N,U_1] & \sqrt{NN'}\,U_2-U_1\sqrt{NN'} \\ \sqrt{NN'}\,U_1-U_2\sqrt{NN'} & [N',U_2] \end{array}\right)\,,
\end{equation*}
and see that that of (a) (\ie $U_1=V$ and $U_2=\Id$) is granted if
\begin{equation}
\label{condcase1}
[N,V]\in\ideal{2}\,,\qquad(\Id-V)\sqrt{NN'}\in\ideal{2}\,;
\end{equation}
and that of (b) ($U_1=V=U_2$) if
\begin{equation}
\label{condcase2}
[N,V]\in\ideal{2}\,,\qquad[\sqrt{NN'}, V]\in\ideal{2}\,.
\end{equation}

We can now complete the proof of parts (ii) and (iii) of the
theorem. For $V=\exp(\iu\lambda Q)$, we use (a). Then eqs.~(\ref{condcase1})
hold true by eqs.~(\ref{QN}, \ref{mixedcond}):
$(\Id-\ep{\iu\lambda Q})\sqrt{NN'}=(1-\ep{\iu\lambda})Q\sqrt{NN'}$.
This also proves (iii). For $V=U$, we use (b), with
conditions~(\ref{condcase2}) holding by eq.~(\ref{NTrCl}, \ref{deltaN}).

Part~(iv) is readily proven as follows. Like in the pure case, the
implementers are unique up to left multiplication by an element of
$\pi_N(\alg{\hilb})'$ (which is larger than the multiples of the identity
since the representation is reducible).
Thus $\exp{(\iu\lambda\widehat{Q})}\in\pi_N(\alg{\hilb})''$ still implies its
uniqueness up to a phase because of (\ref{factor}).\teop\\

\noindent
{\bf Remark.} If it were for property (ii) only, one could adopt method (b)
also for $V=\exp(\iu\lambda Q)$.

\paragraph{Proof of Lemma~\ref{cyclic}.} The space $\hilb_{P_N}$ is spanned by the vectors
\begin{equation}
\label{H_PN}
\prod_{i=1}^n\,\pi_{P_N}(a^{\#}(f_i\oplus g_i))\Omega_{P_N}\,.
\end{equation}
It suffices to show that they can be approximated arbitrarily well by a sum of
such vectors where, however, $g_i=0$. To this end, we first note that
\begin{align*}
\pi_{P_N}(a\str(P_N(f\oplus g)))\Omega_{P_N}=0\,, \\
\pi_{P_N}(a(P_N'(f\oplus g)))\Omega_{P_N}=0\,,
\end{align*}
where $P_N'=1-P_N$. This follows from~(\ref{state}) for $P_N$ instead of $N$, and implies in turn
\begin{align*}
\| \pi_{P_N}(a\str(f\oplus g))\Omega_{P_N}\|\leq\|P_N'(f\oplus g)\|\,, \\
\|\pi_{P_N}(a(f\oplus g))\Omega_{P_N}\|\leq\|P_N(f\oplus g)\|\,.
\end{align*}
Let us first consider the case where the last factor in~(\ref{H_PN}) is an annihilation operator and set $f_n\oplus g_n=:f\oplus g$. We have
\begin{equation*}
P_N(f\oplus g-\widetilde{f}\oplus 0)=\left(\begin{array}{c}
\sqrt{N} \\ \sqrt{N'}
\end{array}\right) \bigl(\sqrt{N}(f-\widetilde{f})+\sqrt{N'}g\bigr)\,.
\end{equation*}
A vector $\widetilde{f}\in\hilb$ is well-defined by
\begin{equation*}
\sqrt{N}\widetilde{f}:=\sqrt{N} f+\sqrt{N'}F(N\geq\epsilon)g\,,
\end{equation*}
where $F(N\geq\epsilon)$ is the spectral projection for $N$ on $[\epsilon,
\infty)$ and $\epsilon>0$. Thus,
\begin{equation*}
\|P_N(f\oplus g-\widetilde{f}\oplus 0)\|\leq\|F(N < \epsilon)g\|
\end{equation*}
can be made arbitrarily small because of $\Ker{N}=\{0\}$. If the last factor is a creation operator, the arguments proceed similarly using $\Ker{N'}=\{0\}$. Hence the announced replacement can be performed in the last factor. After anticommuting it to the left, the claim is reduced to products with fewer factors, for which it holds by induction.\teop

\subsection{Part (v)}

The idea of the proof is to approximate the Bogoliubov automorphism induced by
$\ep{\iu\lambda Q}$ by means of inner automorphisms, as introduced in
Subsection~\ref{pre}. The generating function on the l.h.s. of 
(\ref{Main}) then becomes computable by Lemma~\ref{scalprod}.
We present separate proofs in the pure and the mixed case. The
second proof, while applying to both cases, is longer than the one we
give for pure states. Both depend on Prop.~\ref{continuity}.

{\bf Pure state.}
Let $F$ be a finite rank operator on $\hilb$ with $[F,N]=0$. As such, it has
an implementation in the cyclic representation $\pi_N$; its non-uniqueness
does not affect the l.h.s. of
\begin{align}
(\Omega_N,
\widehat{U}\str\ep{\iu\lambda\widehat{F}}\widehat{U}\ep{-\iu\lambda\widehat{F}}
\Omega_N)
&=(\Omega_N,
\widehat{U}\str\pi_N(\Gamma(\ep{\iu\lambda F}))\widehat{U}
\pi_N(\Gamma(\ep{-\iu\lambda F}))
\Omega_N) \nonumber\\
\label{approxscal}
&=(\Omega_N, \pi_N(\Gamma(U\str\ep{\iu\lambda F}U\ep{-\iu\lambda F}))\Omega_N)\,;
\end{align}
on the r.h.s. we used that $\pi_N(\Gamma(\ep{\iu\lambda F}))$ is one possible
implementation of $\ep{\iu\lambda F}$ by (\ref{prp1}) with
$\ep{\iu\lambda F}$ in place of $U$; the second
line follows by (\ref{UImpl}), which implies
$\widehat{U}\str\pi_N(\Gamma(\ep{\iu\lambda F}))\widehat{U}=
\pi_N(\Gamma(U\str\ep{\iu\lambda F}U))$, and by (\ref{prp3}). Another choice
for $\widehat{F}$ is fixed by
\begin{equation}
(\Omega_N,\widehat{F}\Omega_N)=0\,,
\label{nrmlz1}
\end{equation}
and we may ask the same normalization for $\widehat{Q}$.
\begin{lma}
\label{Fnpure}
There is a sequence of finite dimensional orthogonal projections $F_n$ such
that
\begin{equation}
[F_n,N]=0\,,\qquad\mathrm{s-}\lim_{n}\,F_n=Q\,.
\label{FnN}
\end{equation}
\end{lma}
%
\paragraph{Proof.} We note that $(NQ)^2=NQ$, so that $Q=NQ+N'Q$ is an
orthogonal splitting of $Q$. Let $F_n=F_n^{(1)}+F_n^{(2)}$, where $F_n^{(1)}$,
resp. $F_n^{(2)}$, is a subprojection of $NQ$ (\ie $F_n^{(1)}NQ=F_n^{(1)}$),
resp. of $N'Q$, with $F_n^{(1)}\stackrel{s}{\to} NQ$, and $F_n^{(2)}\stackrel{s}{\to} N'Q$. Clearly, $F_n\stackrel{s}{\to}Q$ and
\begin{equation*}
[F_n^{(1)},N]=[N',F_n^{(1)}]=N'F_n^{(1)}-F_n^{(1)}N'=
N'NQF_n^{(1)}-F_n^{(1)}QNN'=0\,,
\end{equation*}
since $NN'=0$. The same holds for $F_n^{(2)}$, and thus for $F_n$.\teop\newline

By (\ref{FnN}, \ref{nrmlz1}) the assumptions of Prop.~\ref{continuity} are
satisfied for the sequence $(F_n)$ and its limit $Q$. Therefore,
\begin{equation}
\label{scalprodconv}
(\Omega_N, \widehat{U}\str\ep{\iu\lambda\widehat{Q}}\widehat{U}\ep{-\iu\lambda\widehat{Q}} \Omega_N)=\lim_{n\to\infty}(\Omega_N, \widehat{U}\str\ep{\iu\lambda\widehat{F_n}}\widehat{U}\ep{-\iu\lambda\widehat{F_n}} \Omega_N)\,.
\end{equation}
By eq.~(\ref{approxscal}, \ref{det}, \ref{FnN}) the inner product on the
r.h.s. equals
\begin{equation*}
\det(N'+\ep{\iu\lambda U\str F_n U}\ep{-\iu\lambda F_n}N)=
\det(\ep{-\iu\lambda N_U F_{nU}}N'\ep{\iu\lambda N F_{n}}+
\ep{\iu\lambda N'_U F_{nU}}N\ep{-\iu\lambda N' F_n})\,,
\end{equation*}
where we multiplied the determinant by
\begin{equation}
\label{purecasemultdet}
1=\det(\ep{-\iu\lambda N_U F_{nU}})\cdot\det(\ep{\iu\lambda N F_{n}})\,,
\end{equation}
like in the heuristic derivation (\ref{formal}); but unlike there, this step
is now correct, since $F_n$ is of finite rank. We also used $[F_n,N]=0$.
Finally, we claim that the operator under the last determinant converges to
\begin{equation}
\label{needalabel}
\ep{-\iu\lambda N_U Q_U}N'\ep{\iu\lambda N Q}+\ep{\iu\lambda N'_U Q_U}N\ep{-\iu\lambda N' Q}=\ep{-\iu\lambda N_U Q_U}N'+\ep{\iu\lambda N'_U Q_U}N
\end{equation}
in trace class norm, \ie the same expression with $Q$ in place of $F_n$.
The r.h.s. is obtained using $\exp(\iu\lambda
NQ)=\Id+NQ(\exp(\iu\lambda)-1)$ and $NN'=0$. The convergence implies that of
the determinants: Indeed, for $A-\Id$, $B-\Id\in\ideal{1}$, we have (\cite{RSIV}, Lemma~XIII.17.1 (d))
\begin{equation*}
|\det A-\det B|\leq\|A-B\|_1\ep{(\|A-\Id\|_1+\|B-\Id\|_1+1)}\,.
\end{equation*}
Upon conjugating with $U$, it is enough to show
\begin{equation*}
\|(\ep{-\iu\lambda N F_n}-\ep{-\iu\lambda N Q})N'_{U\str}\|_1
\longrightarrow 0\,,
\end{equation*}
and similarly with $N$ and $N'$ interchanged. This operator equals
$\ep{-\iu\lambda}-1$ times
\begin{equation*}
N(F_n-Q)N'_{U\str}=
(F_n-Q)NN'+(F_n-Q)N(N'_{U\str}-N')\,.
\end{equation*}
The first term vanishes, and the second tends to $0$ in the trace class norm
as $n\to\infty$, because of
\begin{equation}\label{trclconv}
X_n\stackrel{s}{\longrightarrow}0\,,\quad Y\in\ideal{1}\quad\Longrightarrow\quad\| X_nY \|_1\longrightarrow 0\;.
\end{equation}
\teop\newline

{\bf Mixed state.}
Let us start by proving a result analogous to Lemma~\ref{Fnpure}:
\begin{lma}
\label{lmahyp0}
Let $P$, $Q$ be orthogonal projections in a separable Hilbert space $\hilb$ with
\begin{equation}
\label{lmahyp}
[Q,P]\in\ideal{1}\,.
\end{equation}
Then there are finite dimensional subprojections $F_n$ of $Q$ with
\begin{equation}
\|[F_n-Q,P]\|_1\longrightarrow 0\,,\quad(n\to\infty)\,.
\end{equation}
\end{lma}

\paragraph{Proof.} We split $Q$ as
\begin{equation}
\label{splitting}
Q=QPQ+Q(1-P)Q\equiv L_1+L_0\,,
\end{equation}
and observe that $[Q,L_1]$=0 and
\begin{gather}
\label{PLone}
(P-1)L_1\in\ideal{1}\,, \\
L_1^2-L_1=QP[Q,P]Q\in\ideal{1}\,.\nonumber
\end{gather}
By the last property, the only possible accumulation points in the spectrum of $L_1$ are $0$ and $1$. In particular, there is an $x\in(0,1)$ which is not in the spectrum. Let $Q_1$ be the spectral projection of $L_1$ associated with $(x,\infty)$. It may be represented as
\begin{equation*}
Q_1=\frac{1}{2\pi\iu}\oint_{\mathcal{C}}(z-L_1)^{-1}dz\,,
\end{equation*}
where $\mathcal{C}\subset\mathbb{C}$ is a contour encircling that part of the spectrum only. Using $\oint_{\mathcal{C}}z^{-1}dz=0$, due to $x>0$, we have
\begin{align}
(P-1)Q_1&=\frac{1}{2\pi\iu}\oint_{\mathcal{C}}(P-1)\left((z-L_1)^{-1}-z^{-1}\right)dz \nonumber \\
\label{PQone}
&=\frac{1}{2\pi\iu}\oint_{\mathcal{C}}(P-1)L_1(z-L_1)^{-1}z^{-1}dz\in\ideal{1}
\end{align}
by~(\ref{PLone}). On the subspace $\Ran Q$, the projection $Q_1$, defined in terms of $L_1$ and $x$ is complementary to the one, $Q_0$, similarly defined by $L_0$ and $1-x$, see~(\ref{splitting}). Since $1-x>0$, we have
\begin{equation}
\label{PQnought}
PQ_0\in\ideal{1}
\end{equation}
by analogy to~(\ref{PQone}). Let now $F_n^{(i)}$, $(i=0,1)$, be a sequence of finite dimensional subprojections of $Q_i$ with $F_n^{(i)}\stackrel{s}{\rightarrow}Q_i$. Then
\begin{align*}
[F_n^{(0)}-Q_0,P]&=(F_n^{(0)}-Q_0)P-P(F_n^{(0)}-Q_0)=(F_n^{(0)}-Q_0)Q_0P-PQ_0(F_n^{(0)}-Q_0)\,, \\
[F_n^{(1)}-Q_1,P]&=[F_n^{(1)}-Q_1,P-1]=(F_n^{(1)}-Q_1)Q_1(P-1)-(P-1)Q_1(F_n^{(1)}-Q_1)\,,
\end{align*}
are trace class by~(\ref{PQone}, \ref{PQnought}), and converge to zero in the corresponding norm by~(\ref{trclconv})
and $\| T\str \|_1=\| T \|_1$. Thus $F_n=F_n^{(0)}+F_n^{(1)}$ is seen to have the stated properties.\teop\newline

We apply the lemma to $\hilb\oplus\hilb$, $P_N$ and $\widetilde{Q}=Q\oplus0$ instead of $\hilb$, $P$ and $Q$; in this case, subprojections of $\widetilde{Q}$ are of the form $F\oplus 0$, with $F$ a subprojection of $Q$. Since
\begin{equation*}
[P_N, \widetilde{Q}]=\left(\begin{array}{cc}[N,Q] & -Q\sqrt{NN'} \\ \sqrt{NN'}Q & 0\end{array}\right)
\end{equation*}
the hypothesis~(\ref{lmahyp}) of Lemma~\ref{lmahyp0} is fulfilled. The
claim yields
\begin{equation}
\label{traceconv}
\|[F_n-Q,N]\|_1\stackrel{n\to\infty}{\longrightarrow}0\,,
\end{equation}
as well as $\|\sqrt{NN'}(F_n-Q)\|_1\rightarrow0$, which however is already
known by~(\ref{mixedcond}) and $F_n=F_nQ$. We thus have a sequence $(F_n)$ of
unitarily implementable transformations: the conditions~(\ref{condcase1}) are
both fulfilled, the first one because $[N,\exp(-\iu\lambda
F_n)]=[N,F_n](\exp(-\iu\lambda)-1)$ and the second because
$(\Id-\exp(-\iu\lambda
F_n))\sqrt{NN'}=(\exp(-\iu\lambda)-1)F_nQ\sqrt{NN'}$. Moreover, the
assumptions of Prop.~\ref{continuity} are satisfied, so that
eqs.~(\ref{scalprodconv},\ref{approxscal}) are true again.

To complete the proof, it remains to show that
\begin{equation}
\label{detConvMix}
\det(N'+\ep{\iu\lambda F_{nU}}\ep{-\iu\lambda F_n}N)\longrightarrow\det(\ep{-\iu\lambda N_U Q_U}N'\ep{\iu\lambda N Q}+\ep{\iu\lambda N'_U Q_U}N\ep{-\iu\lambda N' Q})\,.
\end{equation}
To this end, we multiply the determinant by
\begin{equation}
\label{mixedcasemultdet}
\det(1+F_{nU}(\ep{-\iu\lambda N_U}-1))\,,\quad\det(1+(\ep{\iu\lambda N}-1)F_{n})\,,
\end{equation}
from the left, resp. from the right. These factors would be identical to those in~(\ref{purecasemultdet}) if $F_n$ and $N$ commuted, which is however no longer the case. Also, their product is not $1$, but rather equals
\begin{multline}\label{detcorr}
\det(1+F_{nU}(\ep{-\iu\lambda N_U}-1))\cdot\det(1+(\ep{\iu\lambda N}-1)F_{n})\\
\begin{aligned}
&=\det(1+(\ep{\iu\lambda N}-1)F_n)\cdot\det(1+F_n(\ep{-\iu\lambda N}-1))\\
&=\det(1-F_n+\ep{\iu\lambda N}F_n\ep{-\iu\lambda N})\,,
\end{aligned}
\end{multline}
where
\begin{equation*}
\ep{\iu\lambda N}F_n\ep{-\iu\lambda N}-F_n=\iu\int_0^{\lambda}\,\ep{\iu sN}[N,F_n]\ep{-\iu s N}ds\in\ideal{1}
\end{equation*}
and
\begin{equation}
\label{supp}
\|\ep{\iu\lambda N}F_n\ep{-\iu\lambda N}-F_n\|_1\stackrel{n\to\infty}{\longrightarrow} 0
\end{equation}
by $[N,Q]=0$ and~(\ref{traceconv}). Therefore, (\ref{detcorr}) converges to
$1$ and it suffices to prove~(\ref{detConvMix}) with the l.h.s. multiplied
by~(\ref{mixedcasemultdet}). The determinant becomes that of
\begin{equation}
\label{multipieddet}
(1+F_{nU}(\ep{-\iu\lambda N_U}-1))(N'+\ep{\iu\lambda F_{nU}}\ep{-\iu\lambda F_n}N)(1+(\ep{\iu\lambda N}-1)F_n)\,.
\end{equation}
By means of
\begin{align}
\label{firstconv}
&\|(1+F_{nU}(\ep{-\iu\lambda N_U}-1))N'-\ep{-\iu\lambda N_UQ_U}N'\|_1\longrightarrow 0\,, \\
\label{secconv}
&\|\ep{-\iu\lambda F_n}N\ep{\iu\lambda F_n}-N\|_1\longrightarrow 0\,,
\end{align}
which we shall prove momentarily, we may replace~(\ref{multipieddet}) by
\begin{equation*}
\ep{-\iu\lambda N_UQ_U}N'(1+(\ep{\iu\lambda N}-1)F_n)+(1+F_{nU}(\ep{-\iu\lambda N_U}-1))\ep{\iu\lambda F_{nU}}N\ep{-\iu\lambda F_{n}}(1+(\ep{\iu\lambda N}-1)F_n)\,.
\end{equation*}
The claim then follows from
\begin{align}
\label{thirdconv}
&\|N'(1+(\ep{\iu\lambda N}-1)F_n)-N'\ep{\iu\lambda NQ}\|_1\longrightarrow 0\,, \\
\label{fourthconv}
&\|N\ep{-\iu\lambda F_n}(1+(\ep{\iu\lambda N}-1)F_n)-N\ep{-\iu\lambda N'Q}\|_1\longrightarrow 0\,, \\
\label{fifthconv}
&\|(1+F_{nU}(\ep{-\iu\lambda N_U}-1))\ep{\iu\lambda F_{nU}}N-\ep{\iu\lambda N'_UQ_U}N\|_1\longrightarrow 0\,,
\end{align}
It remains to prove (\ref{firstconv}\,-\,\ref{fifthconv}). The limit~(\ref{secconv}) follows like~(\ref{supp}). The expression in~(\ref{thirdconv}) is $N'(\exp(\iu\lambda N)-1)(F_n-Q)=f(N)\,NN'Q(F_n-Q)$ where $f(N)=N^{-1}(\exp(\iu\lambda N)-1)$ is a bounded operator; its convergence to zero follows from~(\ref{mixedcond}). As for~(\ref{fourthconv}) we have
\begin{align*}
\ep{-\iu\lambda F_n}(1+(\ep{\iu\lambda N}-1)F_n)&=(1+(\ep{-\iu\lambda}-1)F_n)(1+(\ep{\iu\lambda N}-1)F_n) \\
&=1+(\ep{-\iu\lambda}\ep{\iu\lambda N}-1)F_n+(\ep{-\iu\lambda}-1)[F_n,\ep{\iu\lambda N}] F_n\,,
\end{align*}
so that by using~(\ref{supp}) it remains to show
\begin{equation*}
\|N(1+(\ep{-\iu\lambda N'}-1)F_n)-N\ep{-\iu\lambda N'Q}\|_1\longrightarrow 0\,.
\end{equation*}
This, however, is just~(\ref{thirdconv}) with $N$ and $N'$ interchanged. Finally, in~(\ref{firstconv}, \ref{fifthconv}) we may, by~(\ref{NTrCl}), replace $N$ and $N'$ by $N_U$ and $N_U'$ in those places where the subscript is not already present. By passing to a unitary conjugate and adjoint, they reduce to~(\ref{thirdconv}, \ref{fourthconv}).\teop

\section{Examples}\label{exa}
We illustrate the hypotheses~(\ref{QN}, \ref{NTrCl}) by presenting a model in
which they can be verified. The left and right portions of the single lead
mentioned in Sect.~1 are replaced by two infinite leads, which are however
chiral. The interaction between them occurs in a finite interval and allows
particles to scatter between the leads.

\begin{figure}[h]
\begin{center}
\input{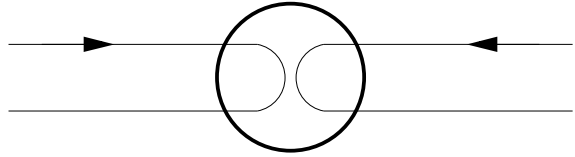}
\caption{A simple model with two infinite chiral leads}\label{fig_chiralPump}
\end{center}
\end{figure}

Let $\hilb=L^2(\mathbb{R})\oplus L^2(\mathbb{R})$ be the one-particle space with operators
\begin{equation}
Q=\left(\begin{array}{cc}
1 & 0 \\ 0 & 0
\end{array}\right)\,,\qquad N=\left(\begin{array}{cc}
\Theta(-p) & 0 \\ 0 & \Theta(p)
\end{array}\right)\,.
\end{equation}
Here, $x\in\mathbb{R}$ is the position variable, $p=-\iu d/dx$ the conjugate
momentum, and $\Theta$ the Heaviside function. The projection $N$ describes
the Fermi sea of the free Hamiltonian
\begin{equation}
\label{ex1_hamiltonian}
H_0=\left(\begin{array}{cc}
p & 0 \\ 0 & -p
\end{array}\right)
\end{equation}
for vanishing Fermi energy. Clearly, $[Q,N]=0$.

\subsection{Example 1}\label{Example 1}

The example is conveniently stated by passing to another pair of
conjugate variables, $E$ and $t$: The energy $E:=\pm p$ yields the 
spectral representation of $H_0$ in multiplication form,
\[
H_0=\left(\begin{array}{cc}
E & 0 \\ 0 & E
\end{array}\right)\,,
\]
while the operator
\[
T=\left(\begin{array}{cc}
-x & 0 \\ 0 & x
\end{array}\right)
=:
\left(\begin{array}{cc}
t & 0 \\ 0 & t
\end{array}\right)\,,
\]
represents, due to $\iu [H_0,T]=-1$, the time $t$ of passage at $x=0$ of
a freely moving particle, which is presently elsewhere. In this 
example only, the meaning of $t$ is
therefore that of a dynamical variable, and not that of the parameter governing
evolution.

Rather than specifying an interacting Hamiltonian, we model the
scattering process by directly giving the propagator $U$ for the
time interval under consideration. We assume it to be given by a
unitary multiplication operator $U(t)$ with $U(t)-1$ of compact
support, see Remark~1 in Sect.~3.

Such a simple kind of evolution should be seen as an effective description in
the adiabatic limit and in the interaction picture. The passage across the
interaction region maps the incoming state to the outgoing one
by means of a scattering matrix which, in the limit of low frequencies
$\omega$, is that of the static scatterer in effect at time $t$, $S(t)$. In
the same limit, only electrons within an interval $\sim\hbar\omega$ of the
Fermi energy ought to matter for transport. Thus, $U(t)=S(t,0)$, where
the $2\times 2$-matrix $S(t,E)$ is the fiber of $S(t)$ at energy $E$. For a
more thorough justification, see~\cite{BPT, AEGS}.
\begin{prop}
\label{exa1}
Suppose $U-1\in M_2(C_0^{\infty}(\mathbb{R}_t))$. Then $[N,U]\in \ideal{1}$.
\end{prop}
Here $M_2(X)$ are the $2\times 2$ matrices with entries in $X$.

\paragraph{Proof.} We may rename $U(t)-1$ by $U(t)$
without loss. By the assumption we may write
$U=fU$, where $f=f(t)$ satisfies $f\in
C_0^{\infty}(\mathbb{R}_t)$, too. Then
$[N,U]=f[N,U]+[N,f]U$, with
\begin{equation*}
f[N,U]=f(E+\iu)^{-1}\cdot(E+\iu)[N,U]\,,
\end{equation*}
and similarly for the second term. We claim that both factors are 
Hilbert-Schmidt and hence
their product trace class. The first one is, because the functions $f$ and
$g(E)\equiv (E+\iu)^{-1}$ are in $L^2(\mathbb{R}_t)$, resp. 
$L^2(\mathbb{R}_E)$. As for the second
one, we note that $N=\Theta(-E)\otimes 1_2$, whence $[N,U]$ has
matrix entries $[\Theta(-E),U_{ij}]$. That leads to integral
operators $K$ acting merely on $L^2(\mathbb{R}_E)$ with kernels
\begin{equation*}
K(E,E')=(E+\iu)\widehat{U}_{ij}(E'-E)\bigl(\Theta(-E)-\Theta(-E')\bigr)\,.
\end{equation*}
They are supported where $\mathrm{sgn}\,E = -\mathrm{sgn}\,E'$ and satisfy
\begin{equation*}
\iint |K(E,E')|^2 dE\,dE'=
\int_0^{\infty}\int_0^{\infty}|E''+\iu|^2\bigl(
|\widehat{U}_{ij}(E'+E'')|^2 +|\widehat{U}_{ij}(-E'-E'')|^2\bigr)dE\,dE''\,,
\end{equation*}
which is finite. Thus, the corresponding operator is in
$\ideal{2}$.\teop\newline

By contrast, but under the same assumption as in the proposition, the
operator $(Q_U-Q)N$ may fail to be trace class. By~(\ref{ExpTransp}), this
shows the need for regularizing~(\ref{Levitov}). Indeed, we may arrange for a
$\psi\in\hilb$ and $U$ such that $(Q_U-Q)\psi\neq 0$. The sequence
$\psi_n=\exp(\iu n T)\psi$ tends to zero weakly. Using
$\Theta(-E)\exp(\iu n t)=\exp(\iu n t)\Theta(n-E)$ and
$\Theta(n-E)\stackrel{s}{\rightarrow}1$, we
have $\|N\psi_n-\psi_n\|\rightarrow 0$ and, since $Q$, $U$ are multiplication
operators in $t$, $\|(Q_U-Q)N\psi_n\|\rightarrow\|(Q_U-Q)\psi\|\neq 0$. As a
result, $(Q_U-Q)N$ is not even compact.
The argument just given may be summarized in physical terms as follows: 
Whatever contribution to transport, as signified by $(Q_U-Q)N$, comes from one
energy in the Fermi sea, it is repeated at all such energies, because the
evolution $U$ proceeds with the same velocity $\pm 1$ at all energies. 

It should be remarked that $[N,U]$ may fail to be in
$\ideal{2}$ if, unlike in Prop.~\ref{exa1}, $U(t)$ attains different
limits at $t\to\pm\infty$. This fact has been pointed out in
\cite{LevitovLeeLesovik} in slightly different terms as a
manifestation of the orthogonality catastrophe. Consider for
instance a potential drop $V(t)$ of finite duration being applied
between the leads, with $\int_{-\infty}^\infty V(t)dt\notin
2\pi\mathbb{Z}$. That situation can be modeled in the context of the
present example by means of a vector potential, where it gives raise
to the catastrophe. The same physical situation is however tame in
the context of the next example.

\subsection{Example 2}

Here we specify a time-dependent perturbation of (\ref{ex1_hamiltonian}),
$H(t)=H_0+V(t)$, where $V(t)$ is multiplication by a $2\times 2$ matrix
$V(t,x)$. Let $U=U(t_2,t_1)$ be the propagator for $H(t)$ between times $t_1$
and $t_2$.
\begin{prop}
Suppose
$V(t,\cdot),\,\partial_t V(t,\cdot)\in M_2(C_0^{\infty}(\mathbb{R}_x))$.
Then $[N,U]\in \ideal{2}$.
\end{prop}
Note that the commutator is claimed to be Hilbert-Schmidt only, which covers
only the statements (ii-iv) of Theorem~\ref{thm1}.
\paragraph{Proof.} By \cite{RSe}, Lemma 4  or \cite{Ru}, Thm. 2.8 it suffices
to show that the statement holds true for the first term in the Dyson
expansion of $U$, \ie for
\begin{equation}
\label{dyson}
\tilde U(s_2,s_1)=-\iu\int_{s_1}^{s_2}\ep{\iu H_0t}V(t)\ep{-\iu H_0t} dt \,,
\end{equation}
with estimates uniform in the sub-interval $[s_1,s_2]\subset[t_1,t_2]$.
By writing
\[
V(t)=\left(\begin{array}{cc}
V_{++}(t)&V_{+-}(t)\\
V_{-+}(t)&V_{--}(t)
\end{array}\right)\,,
\]
the kernel of $[N,V(t)]$ in momentum space becomes
\[
[N,V(t)](p,p')=
\left(\begin{array}{cc}
\hat V_{++}(t,p-p')(\Theta(-p)-\Theta(-p'))&
\hat V_{+-}(t,p-p')(\Theta(-p)-\Theta(p'))\\
\hat V_{-+}(t,p-p')(\Theta(p)-\Theta(-p'))&
\hat V_{--}(t,p-p')(\Theta(p)-\Theta(p'))
\end{array}\right)\,.
\]
The diagonal contributions are in $\ideal{2}$ without recourse to the
integration (\ref{dyson}). For instance,
\[
\iint dpdp'\,|\hat V_{--}(t,p-p')|^2|\Theta(p)-\Theta(p')|
=\int_{-\infty}^\infty du\,|u||\hat V_{--}(t,u)|^2<\infty\,.
\]
The off-diagonal contributions improve once the time integral is performed.
We compute it by parts and obtain, for instance, the kernel
\begin{multline}
\label{pint}
-\iu\int_{s_1}^{s_2}\hat V_{+-}(t,p-p')
\ep{\iu(p+p')t}dt\\
=-\frac{\ep{\iu(p+p')t}-1}{p+p'}\hat V_{+-}(t,p-p')\Big|_{s_1}^{s_2}
+\int_{s_1}^{s_2}\frac{\ep{\iu(p+p')t}-1}{p+p'}
\partial_t \hat V_{+-}(t,p-p')dt\,,
\end{multline}
times $\Theta(-p)-\Theta(p')$. The boundary terms are separately in
$\ideal{2}$, since their corresponding square norm is
\begin{multline*}
4\iint dpdp'\,\frac{\sin^2((p+p')s_i/2)}{(p+p')^2}
|\hat V_{+-}(s_i,p-p')|^2|\Theta(-p)-\Theta(p')|\\
=4\int_{-\infty}^\infty du\,\frac{\sin^2(us_i/2)}{u^2}
\int_{-|u|/2}^{|u|/2}dv\, |\hat V_{+-}(s_i,v)|^2
\le \pi |s_i| \|V_{+-}(s_i)\|_2^2\,.
\end{multline*}
By the same estimate, but with $\partial_t V_{+-}(t)$ in place of
$V_{+-}(s_i)$, also the integrand in (\ref{pint}) is in
$\ideal{2}$. \teop\newline

We recall that in \cite{RSe, Ru} the implementation of the propagator of a
time-dependent Dirac Hamiltonian was studied, of which the above $H(t)$ is the
1-dimensional version. In larger dimensions, as considered there, the
implementability is ensured only in some cases.

We remark that by the method used in Example 1 one can show that
diagonal perturbations lead to $[N,U]\in \ideal{1}$, but not for off-diagonal
ones.

\paragraph{Acknowledgements.} We would like to thank H. Araki,
P. Deift, G. Dell'Antonio, G. Kottanattu, G. Lesovik and W. de Roeck
for discussions.
We also thank the Erwin Schr\"odinger Institute (Vienna) and the
Lewiner Institute for Theoretical Physics at the Technion (Haifa) for
hospitality.


\end{document}